\documentstyle[aps,epsf,epsfig,prl,psfig,preprint]{revtex}

\begin{document}
\draft
\title{Combinatorial and topological approach to the 3D Ising model}
\author{Tullio Regge\cite{tr} and Riccardo Zecchina \cite{rz}}
\address{\cite{tr} Dip. Fisica, Politecnico di Torino, \\
C.so Duca degli Abruzzi 24, I-10129 Torino, Italy \\
\cite{rz} The Abdus Salam International Centre for Theoretical Physics, \\
Strada Costiera 11, P.O. Box 586, 34100 Trieste, Italy}
\maketitle

\begin{abstract}
We extend the planar Pfaffian formalism for the evaluation of the Ising 
partition function to lattices of high topological genus $g$. 
The 3D Ising model on a cubic lattice, where $g$ is proportional to the
number of sites, is discussed in detail. The expansion of the partition
function is given in terms of $2^{2 g}$ Pfaffians classified by the oriented 
homology cycles of the lattice, i.e. by its spin-structures. 
Correct counting is guaranteed by a signature term which depends on the 
topological intersection of the oriented cycles through a simple bilinear 
formula. The role of a gauge symmetry arising in the above expansion is 
discussed.

The same formalism can be applied to the counting problem of perfect
matchings over general lattices and provides a determinant expansion of the
permanent of 0-1 matrices.
\end{abstract}



\section{Introduction}

The evaluation of the matching polynomial of a general graph with weighted
edges is at the same time a root problem for discrete mathematics,
statistical mechanics and mathematical chemistry. Even in its simplest
version, the so called {\it dimer covering problem}, in which the sites of a
lattices have to be covered by non-overlapping arrangements of dimers, the
evaluation of the perfect matching polynomial is a fundamental problem of
lattice statistics\cite{Kast2,Welsh,Domb,McCoy,Lovasz}. For planar graphs,
e.g. $2D$ regular lattices, the counting problem is easily reduced via
Kasteleyn's theorem on lattices orientation to the evaluation of a finite
number of Pfaffians \cite{Kast2,Kast1}. Such a computation requires a number
of operations which is polynomial in the number of vertices and is
considered to be a tractable problem. For instance, the exact analytical
solution of the regular $2D$ Ising model \cite{Onsager} can be easily
obtained by expressing the high temperature loop counting problem in terms
of a dimer covering generating function over a properly decorated lattice 
\cite{Kast2,McCoy,Kast1,Fisher,Hurst}. The periodic nature of Kasteleyn's
orientation allows for the evaluation of the associated Pfaffian by
diagonalization. Similarly, the Pfaffian method has been used in
mathematical chemistry \cite{Chem} to derive the asymptotic number of dimer
coverings for any regular surface lattice. Such a number is strictly related
to the efficiency of adsorption processes of dimer molecules over surfaces,
or to the degeneracy of double bond arrangements in planar organic lattices
(the so called Kekul\`{e} structures).

In the case of non-homogeneous planar lattices, though the closed-form
analytical solution is in general impossible to obtain, the Ising and the
Dimer problems remain tractable in algorithmic sense \cite{Kardar}.

The nature of the matching problem changes completely if one considers
non-planar graphs or lattices\cite{Lovasz}. In discrete mathematics, it is
known that the counting problem becomes $\# P$-complete \cite{Valiant} and
no exact polynomial algorithm exists for the enumeration of coverings.

In statistical mechanics and mathematical chemistry, the interest in
non-planar lattices hinges on the fact that they are equivalent to
higher dimensional lattices. The $3D$ cubic lattice can be considered as a
handlebody $2D$ lattice of topological genus $g=1+N/4$ where $N$ is the
number of sites. A non vanishing ratio $g/N$ for $N \to \infty$ 
is related to an effective
dimension $D>2$ of the lattice, at least as far as its computational
complexity is concerned. No exact solution exists for any non-planar lattice
model, the simplest case being two coupled $2D$ Ising models. Similarly, no
exact evaluation of dimer coverings over non-planar lattices is available.
Of course, there exist several powerful probabilistic algorithms and
approximate theories which provide quite accurate information, however the
issue of understanding the onset of intractability is a basic open one.

In this work we give an explicit formalism which generalizes Kasteleyn's
method to arbitrary non-planar graphs. A first step in this direction was
obtained in ref. \cite{nosotros} in which the complete solution for the
Ising model on a highly symmetric finite lattice of genus $g=3$ and $N=168$
vertices was presented. Here we shall extend such formalism to any lattice
and provide a general algorithmic procedure for the $3D$ cubic lattice.
Scope of the paper is to link the combinatorial Pfaffian representation used
for planar lattices with the topological features of non planar lattices. As
a result we find an expansion for the 3D partition function in which the
role of spin variables is played by a smaller set of binary topological
excitations describing spin structures of the embedding surface of the
lattice.

Already in 1963, Kasteleyn \cite{Kast2,Kast1} noticed that the matching
polynomial and the Ising partition function could be written as a weighted
sum of $2^{2g}$ Pfaffians. In particular, since that time it is known that
each Pfaffian can be associated to an element of the group 
$({\bf Z}_2)^g\times ({\bf Z}_2)^g$.

In what follows we show that the Ising partition function can be written
as $Z=(2\cosh (\beta J))^{3 N} {\bf Z}_0(X)$,
where $J$ is the spin-spin interaction energy, $X=\tanh(\beta J)$ is the
activity of a bond at inverse temperature $\beta$ and ${\bf Z}_0(X)$ is the
dimer covering generating function given as a series of Pfaffians with a
topological signature. The final formula we shall prove is 
\begin{equation}
{\bf Z}_0(X)= \frac 1{2^g}\sum_{\{{\bf e}_k=\,\,0,1\}} (-1)^{\sum_{k=2}^{2g}
\sum_{k^{\prime }=1}^{k-1}I[{\bf \omega }_k, {\bf \omega }_{k^{\prime }}]\;
e_ke_{k^{\prime }}}\,\,\, {\bf Pf}(\Phi(\sum_{k=1}^{2g}\;e_k\;{\bf \omega }
_k),X) \; .
\end{equation}
where the variables $\{ e_k=0,1 \}$ encode the orientation of the $2 g$ 
elementary homology  cycles, $I[ {\bf \omega }_k,{\bf \omega }_{k^{\prime }}]$
is the topological intersection  matrix of the homology cycles $\omega_k$ and 
$\Phi$ represents the orientation of the lattice.

The paper is organized as follows. In Sec.{\bf II} we outline some basic
results concerning the combinatorial approaches to the $2D$ Ising model and
we briefly remind the main steps of the so called Pfaffian method. In Sec.
{\bf III} we give a thorough description of the topology of the $3D$ cubic
lattice, thereby fixing the notation. Sec.{\bf IV} is devoted to the
generalization of Kasteleyn's theorem and to the description of the gauge
symmetry that such a generalization introduces in the problem. In Sec.{\bf V}
we analyze the set of cycles an co-cycles in terms of which the partition
function will be expressed. The construction of a topological intersection
formula which gives the sign of Pfaffians in the expansion of the partition
function is given in Sec.{\bf VI}. The final constructive procedure is then
presented in Sec.{\bf VII}. Sec.{\bf VIII} contains some preliminary results
on the Pfaffian expansion whereas in Sec.{\bf IX} we discuss the application
of the formalism to the Dimer Covering and the Permanent problems.

Throughout the paper a few numerical results will be given in order to 
provide some (very preliminary) physical insight. 
The analysis of the physical consequences of the formalism together
with the discussion of the technicalities involved will be the subject of
another paper \cite{preparation}.

Independently, in refs. \cite{Altri} some general results that partially 
overlap with ours are proposed.

\section{Review of combinatorial methods}

Despite the fact that the original Onsager solution to the $2D$ Ising model
relied on the algebraic Transfer Matrix Method\cite{Onsager}, the
combinatorial solutions which have followed provide a more direct geometrical
insight into $2D$ critical phenomena and field theories.

While the transfer matrix method can be defined in any dimension, the 
combinatorial approaches strongly depend on the
topology of the space where the lattice is immersed. Very schematically,
in $2D$ the sum over spin configurations is recast as a sum over closed
curves (loops). Such curves are endowed with both an intrinsic topology and
with the extrinsic one of ${\bf R}^2$. Since the Ising action depends only
on the extrinsic geometry of loops, one has to avoid double counting and a
proper cancellation mechanism, a topological term, has to be introduced in
the sum. Such an approach has been developed by Kac and Ward \cite{Kac} and
provides probably the most natural way of taking the continuum limit toward
a field theoretical analysis\cite{Feynman,Itzykson}.

In $3D$, the generalization of the above method encounters enormous
difficulties due to the variety of intrinsic topologies of surfaces immersed
in $3D$ lattices.
Despite the deep work done in the attempt of recasting the critical $3D$ Ising
problem as a string theory \cite{Polyakov}, the problem remains unsolved under
many aspects.

Here we generalize the $2D$ (planar) Pfaffian or Dimer Covering approach to 
the Ising  model, a purely combinatorial and basic tool of discrete mathematics
that has many applications in counting problems \cite{Lovasz}.

In $2D$, this approach relies on the equivalence between loop counting and
dimer coverings (also referred to as {\it perfect matchings}) over a
suitably decorated lattice. Once such a relationship is established the
Pfaffian methods turns out to be simple both for the derivation of exact
solutions (in the cases of periodic lattices) and for the definition of
polynomial algorithms on $2D$ heterogeneous models \cite{McCoy,Kardar}.

Let us briefly remind how the method works in the $2D$ case.
The interaction energy of the Ising model on a planar square lattice $
\Lambda _{2D}$ is given by 
\begin{equation}
H=-J_1\sum_{j=1}^{N_1}\sum_{k=1}^{N_2}\sigma _{j,k}\sigma
_{j,k+1}-J_2\sum_{j=1}^{N_1}\sum_{k=1}^{N_2}\sigma _{j+1,k}\sigma _{j,k}
\label{Hamiltonian}
\end{equation}
where $N_1,N_2$ are the number of sites in the two orthogonal directions, $
J_1,J_2$ are the spin-spin interaction energies and $\sigma _{j,k}=\pm 1$.
The partition function $Z=\sum_{\{\sigma =\pm 1\}}\exp (-\beta H)$ can be
written as 
\begin{eqnarray}
Z &=&(\cosh (\beta J_1)\cosh (\beta J_2))^{N_1N_2}\sum_{\{\sigma =\pm
1\}}\left[ \prod_{j=1}^{N_1}\prod_{k=1}^{N_2}(1+X_1\sigma _{j,k}\sigma
_{j,k+1})\right] \times  \nonumber \\
&&\left[ \prod_{j=1}^{N_1}\prod_{k=1}^{N_2}(1+X_2\sigma _{j,k}\sigma
_{j,k+1})\right]  \label{Z2d}
\end{eqnarray}
where $X_i=\tanh (\beta J_i)$ are called activities of the bonds. Expanding
the product and evaluating the sum over $\{\sigma =\pm 1\}$, all the terms
containing odd powers of $\sigma $'s give no contribution whereas all even
powers may be replaced by $1$. It follows that the partition function
acquires a clear interpretation as generating functions of closed loops with 
$p$ horizontal and $q$ vertical bonds with no overlapping sides. In fact 
denoting with $N_{pq}$ the number of such loops, we have 
\begin{equation}
Z=(2\cosh (\beta J_1)\cosh (\beta J_2))^{N_1N_2}\sum_{p,q}N_{pq}X_1^pX_2^q\;,
\label{generating}
\end{equation}

In turn, the above expansion can be mapped onto the problem of evaluating
the generating function of dimer coverings (the so called weighted matching
polynomial) over a new ``counting'' lattice $\Lambda _{2D}^{\#}$ obtained by
substituting each site of the original lattice with a cluster of six sites
(two triangles with a joining bond) and by assigning activity $1$ to the new
decorating bonds while retaining the activity of the bonds inherited by the
original lattice. The (eight) possible configurations of loop bonds at any
Ising site are in $1-1$ correspondence with perfect dimer configurations on
the decorating cluster. Therefore the sum in (\ref{generating}) coincides with
the generating functions of perfect matching over the decorated lattice.

Finally, in order to compute $Z$ we orient the lattice according to the
Kasteleyn prescription by assigning arrows to each bond in such a way that for
any closed circuit $\ell $ on $\Lambda _{2D}^{\#}$, the number of bonds of
${\cal \ell }$ oriented clockwise is of opposite parity to the number of sites
enclosed by ${\cal \ell }$. The Kasteleyn rules define completely the
orientation for planar lattices, whereas for non-planar lattices , i.e.
lattices which can be immersed on a surfaces of non-trivial topological genus,
we need further sign fixing for loops not homologically trivial (i.e. without
an interior). The dimer covering generating function can then be expressed as
a weighted sum of Pfaffians of the antisymmetric adjacency matrix with
elements given by the activities of the bonds and signs determined by their
orientation. In virtue of Cayley theorem, Pfaffians are computed as square
roots of the determinant of such matrices.
Thus, the Ising partition function can be written explicitly as a determinant
which for uniform interaction energies can be further
block diagonalized by Fourier Transform. The final calculation of a 
$6$ by $6$ determinant leads to the exact closed form expression of the 
$2D$ Ising partition function. 
A thorough discussion of the above procedure can be found in \cite{McCoy}.

Below we shall concentrate on the generalization of the above
construction to the cubic $3D$ lattice. The procedure is however general and
can be straightforwardly generalized to any non planar lattice. A first
explicit example was presented in \cite{nosotros} for the case of group
lattices with non-trivial topological genus. 
The same inductive reasoning used in ref. \cite{nosotros} leads 
to a simple topological expression for the coefficients in the 
Pfaffian expansion.

\section{The 3D Cubic Lattice and Embedding Surface}

We consider $3D$ cubic lattices $\Lambda$ of sides $N_1,N_2,\,N_3$ with $
N=N_1N_2N_3$ sites and periodic boundary conditions. Each vertex $V$ is
identified by a triple of periodic coordinates $\{n_1,n_2,n_3\},
\;n_i=0...N_i-1$ with $V(n_1,n_2,n_3)\equiv V({\rm mod}(n_1,N_1),{\rm mod}
(n_2,N_2),{\rm mod}(n_3,N_3))$. The sites can also be labeled in sequential
order by the single index $q\equiv q(n_1,n_2,n_3)={\rm mod}(n_1,N_1)+N_1{\rm 
mod}(n_2,N_2)+N_1N_2{\rm mod}(n_3,N_3)$ with the inverse relations, $n_1= 
{\rm mod}(q,N_1)$, $n_2={\rm mod}(\frac{q-n_1}{N_1},N_2)$ and $n_3={\rm mod}
( \frac{q-\,n_1-\,n_2N_1}{N_1N_2},N_3)$. In what follows notations and
operations over the integers $n_1,n_2,n_3$ have to be understood modulo $
N_1,N_2,N_3$ respectively.

The lattice $\Lambda $ is invariant under translations $D_i:n_i\to n_i+1$.

The set of $N_b=3 N$ bonds $L_i(q)$, $i=1,2,3$ of $\Lambda $ connects
couples of neighboring sites $\{V(q),V(D_iq)\}$, thus defining the adjacency
or incidence matrix $A$ of $\Lambda $, $A_{q,q^{\prime }}=1$ if $q$ and $
q^{\prime }$ are connected by a bond and $A_{q,q^{\prime }}=0$ otherwise.

We call {\it plaquette} a square face $F_{i_1}(n_1,n_2,n_3)\equiv F_{i_1}(q) 
$ of $\Lambda $ identified by the sequence of vertices, $
V(q),V(D_{i_2}q),V(D_{i_2}D_{i_3}q),V(D_{i_3}q)$, where with the notation $
i_1,i_2,i_3$ we denote a generic cyclic permutation of the indices $1,2,3$. $
\Lambda $ contains $3$ classes of $N$ plaquettes $F_i(q)$, orthogonal the
axes $i$, $(i=1,2,3)$.

The parity of a site is given by $p(q)=(-1)^{n_1+n_2+n_3}$. $\Lambda $
is a bipartite lattice in that edges always connect vertices of opposite
parity.

In order to implement the dimer method we construct an orientable surface $
\Sigma$ without boundary, which contains all the sites and bonds of $\Lambda$
and is the union of a subset of square plaquettes of $\Lambda$. The number $
N_f$ of such plaquettes is $N_f=N_b/2=3N/2$, each bond belonging to two
plaquettes of the surface and each plaquette containing four bonds. It
follows that $N$ and at least one of the numbers $N_1,N_2,N_3$ need to be
even. For simplicity we shall assume $N_i=2M_i$, so that $N=8M$ with $
M=M_1M_2M_3$.

As we shall see, all the above conditions can be matched by a definition of $
\Sigma$ which preserves part of the symmetries of the original lattice.

The topological genus $g$ of the surface, evaluated by Euler's formula, is $
N-N_b+N_f=2(1-g)$, from which it follows $g=1+2M=1+N/4$.

The definition of $\Sigma$ requires that a plaquette $F_{i_1}(q)$ belongs to 
$\Sigma$ only if $n_{i_2}+n_{i_3}$ is odd, and we shall call {\it faces }
such plaquettes. The final result for $\Sigma $ consists in a square beams
periodic structure shown in Fig. 1.

\subsection{Combinatorial Topology of the 3D cubic lattice}

In order to proceed in the generalization of the Pfaffian method it is
useful to recall some basic notions of combinatorial topology \cite{topology}.

Sites, bonds and faces of $\Sigma $ generate Abelian groups, additive modulo 
$2$, of non-oriented chains $C_k(\Sigma ,{\bf Z_2})$ of dimension $k=0,1,2$
respectively. For any $c\in C_k(\Sigma ,{\bf Z_2})$, we have $c+c=0$ , where 
$0$ is the identity of the group.

The linear boundary operator $\delta$ maps chains of different dimensions $
\delta : C_k(\Sigma ,{\bf Z_2})\to C_{k-1}(\Sigma , \; {\bf Z_2})$ and is
defined as follows 
\begin{eqnarray}
\delta V(q) &=&1\;\;\;,  \nonumber \\
\delta L_i(q) &=&V(q)+V(D_iq)\;\;\;,  \nonumber \\
\delta F_{i_1}(q)
&=&L_{i_2}(q)+L_{i_3}(D_{i_2}q)+L_{i_2}(D_{i_3}q)+L_{i_3}(q)\;\;\;.
\label{def_delta}
\end{eqnarray}
Clearly, $\delta ^2=0$. A chain $c_k$ in the kernel of $\delta $, i.e. such
that $\delta c_k=0$, is said to be closed . A {\it circuit} is a closed
sequence of vertices connected by edges, each edge is an element of $
C_1(\Sigma ,{\bf Z_2})$, the sum of edges of a circuit is a closed $1$-chain 
$c_1$.

A chain $c_{k-1}=\delta c_k\in C_k(\Sigma ,{\bf Z_2})$ is called {\it 
boundary}. Since $\delta ^2=0$, a boundary $c_1$ is necessarily closed and
reduces to a sum of circuits.

We will keep the definition of $\delta F_i(q)$ as given by (\ref{def_delta})
also in the improper case in which $F_i(q)$ is not a face and hence $\delta
F_i(q)$ is not a boundary but only a circuit on $\Sigma$.

Boundaries and closed chains of $C_k(\Sigma,{\bf Z_2})$ generate the
subgroups $B_k(\Sigma,{\bf Z_2}) \subset Z_{k}(\Sigma,{\bf Z_2}) \subset
C_k(\Sigma,{\bf Z_2})$, respectively.

The homology group is thus defined by $H_k(\Sigma ,{\bf Z_2})=Z_k(\Sigma , 
{\bf Z_2})/B_k(\Sigma ,{\bf Z_2})$ so that there exists a projection $\pi $
from closed chains to the elements of $H_1$, $\pi :Z_1(\Sigma ,{\bf Z_2})\to
H_1(\Sigma ,{\bf Z_2})$. Boundaries are topologically trivial since they are
mapped by $\pi $ onto the identity $0$ of the homology groups. The elements
of $H_1(\Sigma ,{\bf Z_2})$ are called cycles and the homology group is
generated by a base of $2g$ elementary cycles which are equivalence classes
of closed chains under the addition of boundaries. A closed chain $c_k\;\in
\;Z_k(\Sigma ,{\bf Z_2})$ such that $\pi c_k\;=\gamma _k\;\in H_k(\Sigma , 
{\bf Z_2})$ is called a representative of $\gamma_k.$

The multiplicative functionals on the elements of $C_k(\Sigma ,{\bf Z_2})$
and $H_k(\Sigma ,{\bf Z_2})$ with values $\pm 1$ constitute groups of
co-chains and co-cycles denoted by $C^k(\Sigma ,{\bf Z_2})$ and $H^k(\Sigma
, {\bf Z_2})$ respectively. The definitions of $H_k(\Sigma ,{\bf Z_2})$ and $
H^k(\Sigma ,{\bf Z_2})$ are independent from the triangulation of the surface $
\Sigma $ and in particular will be valid for the decorated lattice $\Gamma $
defined in the following section. The properties of the groups $H_k(\Sigma , 
{\bf Z_2})$ and $H^k(\Sigma ,{\bf Z_2})$ depend only on the topological
features of $\Sigma $ and not on the choice of the tessellating lattice. 

The symmetry properties of $\Sigma$ can be seen more clearly by considering its
embedding into the subset or box $\Xi$ of ${\bf R}^3 $ defined by 
$0\leq x < N_1$, $0\leq y < N_2$, $0\leq z < N_3$ with periodic boundary
conditions.
The complement $\Xi -\Sigma $ is the union of two open and disjoint subsets $
\Xi _{+},\Xi _{-}$ congruent under the translation $S=D_1 D_2 D_3$ and
invariant under $T_i=D_i^2$. $S$ and $\{T_i\}$ generate the symmetry group of 
$\Sigma$ while $D_i$ alone is not a symmetry of $\Sigma $. The closure of 
$\Xi_{+},\Xi _{-}$ is given by $\Xi ^{+}=\Xi _{+}\cup \Sigma ,\; \Xi
^{-}=\Xi_{-}\cup \; \Sigma $ and $\Xi ^{+}\cap \Xi ^{-}\;=\;\Sigma $. We may
conventionally regard $\Xi ^{+}$ as the interior of $\Sigma$. On $\Xi^{+}$,
the bond $L_i(q)$ is convex/concave depending on whether $n_i(q)$ is
odd/even, hence, moving along $\Xi ^{+}$ with $D_i$, we encounter
alternatively convex and concave bonds. On $\Xi ^{+}$, each face has $2$
concave and $2$ convex bonds which interchange if we focus on $\Xi ^{-}$
rather than $\Xi ^{+}$.

\section{Kasteleyn's orientation and Gauge Symmetry}

\subsection{Decorated counting lattice}
In order to write the Ising partition function as a dimer covering
generating function we first construct a decoration $\Gamma$ of $\Sigma$
obtained, following Fisher's prescription \cite{Fisher}, by replacing each
site of coordination $q$ with a graph of $3(q-2)$ points and $q-2$
triangular faces, as shown in Fig. 2 (for the case of interest here $q=6$).
In our case, the decorated surface $\Gamma $ contains $12N$ vertices.

Each face of $\Sigma$ maps into a face of $\Gamma $ composed of $4$ bonds
inherited from $\Sigma $ plus $8$ bonds introduced by the decoration. Each
site of $\Sigma $ originates $4$ triangular faces in $\Gamma $, Fig. 2B. As
we shall see, the above decoration is also equivalent (with respect
to dimer covering configurations)  to a locally non
planar one in which each site of $\Sigma $ yields a complete graph of $6$
vertices, Fig. 2C.

Clearly, the homology groups of $\Gamma $ and $\Sigma $ coincide and will be
identified. 

\subsection{Proper orientations}

The orientation over bonds is an additional geometrical structure $\Phi $
defined by a function $\phi _j(q)=\pm 1$, $j=-3,-2,-1,1,2,3$ such that $\phi
_j(q)=1$ (resp. $-1$) corresponds to a bond oriented from $V(q)$ to $V(D_jq)$
(resp. from $V(D_jq)$ to $V(q)$), and $\phi_{-j}(q)=-D_j^{-1}\phi_j(q)$.

An orientation of the bonds of $\Gamma$ or $\Sigma $ is said to be {\it 
proper } if it satisfies the condition of Kasteleyn's theorem for planar
lattices, i.e. if by moving anticlockwise along the perimeter of a face we
encounter an odd number of oppositely oriented bonds. Unless otherwise
specified we assume in the following that all orientations $\Phi$ are
proper. Consider now a boundary on $\Gamma$ or $\Sigma$ consisting of a
single circuit and containing in the interior $k$ points. By induction it
follows that moving along it anticlockwise we encounter a number of
oppositely oriented bonds with parity opposite to that of $k$. A proper
orientation of $\Sigma$ defines  a proper one for $\Gamma$ as
follows. Bonds of $\Gamma$ inherited from $\Sigma$ are given the same $\phi
_j(q)$. Bonds forming the boundary of an ornating triangle are then all
clockwise oriented so they appear anticlockwise in the adjacent faces of $
\Gamma$ inherited from $\Sigma $ and lead to a proper orientation. Sites of $
\Sigma$ and corresponding decorating clusters of $\Gamma$ fall into $2$
subsets of even and odd sites according to their parity $p(q)$.

\subsection{Gauge group}
Given a proper orientation $\Phi$ of $\Gamma$, it is possible to derive
different but equivalent ones by reversing all the bonds of $\Gamma$ which
are incident to a given site $V(q)$, according to the {\it gauge} operation
(see Fig. 3)

\begin{equation}
g(q){\rm :}\phi _j(q)\to -\phi _j(q)\,\,\,\,,\,\,\forall j  \label{gauge}
\end{equation}

The operations $\{g(q)\}$ generate the {\it gauge} group ${\cal G}$. The
orientations generated by ${\cal G}$ are equivalent and will be identified.
By using ${\cal G}$ we can fix arrows on $\Gamma$ such that they
relate even and odd clusters by a mirror reflection and reversal of
orientations of all the bonds.

\subsection{Generalized Kasteleyn rule for dimer coverings}
A dimer covering of $\Gamma $ is defined by a $2$-coloring (say black or
white) assignment to the edges of $\Gamma $ such that each site of $\Gamma$
belongs exactly to one black bond. By superposing two dimer coverings we
obtain a set of closed black circuits which cover all vertices of $\Gamma$.
By moving around a circuit we encounter an alternating sequence of black
bonds from each of the $2$ coverings, therefore the length of the circuit
must be even. If the circuit is a boundary it must enclose points which are
themselves connected by bonds forming black circuits, thus the number
of points inside must also be even and by moving around the circuit in
whatever direction we encounter an odd number of opposite arrows.

Dimer configurations are in one to one correspondence with terms of the
Pfaffian of an antisymmetric incidence matrix $M(\Phi ,X)$ which will be
uniquely defined further on by the pair $(\Gamma ,\Phi )$. In what follows
we shall use $1$-chains composed by black circuits only, originating from the
superposition of dimer coverings. For simplicity, since no
confusion may arise, we use for the black subgroups and the corresponding
dual groups the same notation $B_1(\Sigma ,{\bf Z_2}),Z_1(\Sigma ,{\bf Z_2})$
and $H_1(\Sigma ,{\bf Z_2})=Z_1(\Sigma ,{\bf Z_2})/B_1(\Sigma ,{\bf Z_2})$
introduced in the previous section for the full groups.

Adding a boundary to a black circuit $c$ does not alter its orientation. The
latter depends therefore only on the image $\gamma =\pi (c)\in H_1(\Sigma , 
{\bf Z_2})$ and defines a functional $\Phi (c)$. From the bond orientation
function $\phi _i(q)$ we define the orientation function $\Phi _{i_1}(q)$
for the boundary of the faces $\delta F_{i_1}(q)$ as the multiplicative and
gauge invariant functional 
\begin{equation}
\Phi _{i_1}(q)=-\phi _{i_2}(q) \phi _{i_3}(D_{i_2}q) \phi
_{-i_2}(D_{i_2}D_{i_3}q) \phi _{-i_3}(D_{i_3}q) \; \;.  \label{face}
\end{equation}

For a black boundary, Kasteleyn's rule implies 
\begin{equation}
\Phi _i(q)=1,\;i=1..3,q=1...N \; \; .  \label{Kast}
\end{equation}

Given an anticlockwise sequence of bonds $L_{k_j}(q_j)$ forming the circuit $
c_1\in C_1(\Sigma ,{\bf Z_2})$ and connecting the points $V(q_j)$, $j=1,...m 
$, the orientation $\Phi (c_1)$ over $H_1(\Sigma ,{\bf Z_2})$ is defined by 
\begin{equation}
\Phi (c_1)=-\Pi _{j=1}^m\phi _{k_j}(q_j) \;\;.  \label{circuit}
\end{equation}

On a lattice of non trivial genus $g$ ($g>0$) such a functional is not an
element of $H^1(\Sigma ,{\bf Z_2})$, i.e. in general $\Phi (c_1+c_2)\neq \Phi
(c_1)\Phi (c_2)$\ , as it can be readily verified in the simple $2D$ Ising
lattice with periodic boundary conditions but also on other finite lattices
\cite{nosotros}. The argument runs as follows.

The single black circuit $c_{1,2}$ given by the sum $c_1+c_2$ of two black
circuits $c_1$ and $c_2$ of parity $\Phi (c_1)$ and $\Phi (c_2)$
respectively, which intersect $p$ times, has parity 
\begin{equation}
\Phi (c_{1,2})=(-1)^p\Phi (c_1)\Phi (c_2) \; \;.  \label{twocircuits}
\end{equation}

This fact will turn out to be essential in determining the signs with which
the Pfaffians have to appear in the expansion of the generating function.
Let us consider the specific example of the sum of two black circuits $c_1$
and $c_2$ of parity $\Phi (c_1)$ and $\Phi (c_2)$ with representatives $
a+\ell$ for $c_1,$ $b+\ell$ for $c_2$ where $\ell$ is a common bond and $a,b$
are paths closed by $\ell$. A representative for the product $c_{12}$ is now 
$a+b$.

We move along $c_1$ in a specific direction encountering $n_a$ opposite
arrows along $a$ and $r=0,1$ opposite arrow coming from $\ell$ with a total
of $n_a+r$. Similarly we proceed along $c_2$ in such a way that when we move
along $c_{1,2}$, $a$ and $b$ are run in the same direction and meet a number 
$n_b+1-r$ of opposite arrows since $\ell$ is run in the opposite direction
in $c_2$ with respect to $c_1$. The total number of opposite arrows along $
c_{1,2}$ is now $n_a+n_b$ while the total on $c_1,c_2$ is $n_a+n_b+1$. This
result is equivalent to (\ref{twocircuits}) for $p=1$ and can be generalized
to multiple intersections.

\subsection{Gauge tree, spin structures and Pfaffian matrices}

The gauge symmetry $G$ can be used to fix the orientation on a subset of
bonds forming a spanning tree ${\cal T}$ in $\Gamma $, see Fig. 4. It will
be seen that both the gauge choice and Kasteleyn's rules lead to a
number of orientations which is $2^{2g}$. Such orientations break all
symmetries of $\Sigma $ and are in one to one correspondence with the $
2^{2g} $ spin structures of a surface of genus $g$.

For each element of the base of $H_1(\Sigma ,{\bf Z_2})$ there exist two
possible orientations, for a total of $2^{2g}$ different global orientations
of $\Sigma $. Given $\Phi$ for the cycles of the homology base, it is
possible to derive the parity of any circuit by the intersection formalism
related to (\ref{twocircuits}).

$\Phi$ defines (modulo gauge transformations), an antisymmetric matrix $
M(\Phi ,X)$ of dimension $12N\times 12N$ with elements labeled by the sites
of $\Gamma $. Changing the gauge produces a new matrix $M^{\prime }(\Phi
,X)=KM(\Phi ,X)K^{-1}$, with $K$ diagonal, which has the same Pfaffian of $
M(\Phi)$.

If $q,q^{\prime }$ are not connected we set $M_{q,q^{\prime }}(\Phi ,X)=0$.
If $q,q^{\prime }$ are connected by a decorating bond $L_i(q)\;$, we have $
M_{q,q^{\prime }}(\Phi ,X)=\phi _i(q)=\pm 1$ depending on the orientation of
the bond. If $q,q^{\prime }$ are connected by a bond inherited from $\Lambda 
$, we set $M_{q,q^{\prime }}(\Phi ,X)=\phi _i(q)X\;=\pm X$ where $X$ is the
so called activity of the bond, $X=\tanh(\beta J)$ with $\beta$ inverse
temperature and $J$ spin interaction energy. The diagonal $12\times 12$
blocks of $M(\Phi ,X)$ describe the decorating clusters displayed in Fig. 2B
and are given by 
\begin{equation}
\left| 
\begin{array}{llllllllllll}
0 & 0 & 0 & 0 & 0 & 1 & -1 & 0 & 0 & 0 & 0 & 0 \\ 
0 & 0 & 0 & 1 & 0 & 0 & 0 & -1 & 0 & 0 & 0 & 0 \\ 
0 & 0 & 0 & 0 & 1 & 0 & 0 & 0 & -1 & 0 & 0 & 0 \\ 
0 & -1 & 0 & 0 & 0 & 0 & 0 & 1 & 0 & 0 & 0 & 0 \\ 
0 & 0 & -1 & 0 & 0 & 0 & 0 & 0 & 1 & 0 & 0 & 0 \\ 
-1 & 0 & 0 & 0 & 0 & 0 & 1 & 0 & 0 & 0 & 0 & 0 \\ 
1 & 0 & 0 & 0 & 0 & -1 & 0 & 0 & 0 & 1 & 0 & 0 \\ 
0 & 1 & 0 & -1 & 0 & 0 & 0 & 0 & 0 & 0 & 1 & 0 \\ 
0 & 0 & 1 & 0 & -1 & 0 & 0 & 0 & 0 & 0 & 0 & 1 \\ 
0 & 0 & 0 & 0 & 0 & 0 & -1 & 0 & 0 & 0 & 1 & -1 \\ 
0 & 0 & 0 & 0 & 0 & 0 & 0 & -1 & 0 & -1 & 0 & 1 \\ 
0 & 0 & 0 & 0 & 0 & 0 & 0 & 0 & -1 & 1 & -1 & 0
\end{array}
\right|  \label{Block}
\end{equation}

We have thus defined all the elements of $M(\Phi ,X)$ which can be written
as 
\begin{equation}
M(\Phi ,X)=\;A+X\;B(\Phi )\;\;  
\label{EmmeAB}
\end{equation}
where $A$ is block diagonal consisting of $N/2$ fixed blocks as in (\ref
{Block}) corresponding to even sites and $N/2$ opposite blocks corresponding
to odd sites. $B(\Phi )$ contains the $\phi _i(q)$. For computational
purposes the matrix $M(\Phi ,X)$ can be replaced by an equivalent one of
dimensions $12M\times 12M$ by a folding procedure which we can sketch as
follows. In this section ${\bf I}_n$ denotes the identity matrix of rank 
$n$.

{\bf i)} The $12N$ sites are grouped into $2$ subsets each of $6N$ sites .
The first subset $Ext$ includes the internal sites of each of the $N$
decorating clusters,  which label in (\ref{Block}) rows and columns $6,...,12$.
Each of these rows and columns have $3$ non zero entries. The second subset $
Int$ contains sites with $2$ decorating bonds (the $2$ non zero entries in ( 
\ref{Block})) and one inherited from $\Sigma $ which appears in $B$.
Schematically $M(\Phi ,X)$\ can be written in blocks \ $M_{ik}$ of the form 
\begin{equation}
M(\Phi ,X)=\left| 
\begin{array}{ll}
M_{11} & M_{12} \\ 
M_{21} & M_{22}
\end{array}
\right| 
\end{equation}
where $i,k=1,2$ label the subsets $Ext,Int$ respectively. 
But now $M_{11}$ is a block diagonal matrix with $Det(M_{11})=1$ and which can
be easily inverted. We write then
\begin{equation}
\det(M)=\det(M_{11})\det(M_{22}-M_{21} M_{11}^{-1} M_{12})=\det(M_a) \; \;.
\label{NonFaUnaPiega}
\end{equation}

$M_a=M_{22}-M_{21}\;M_{11}^{-1}\;M_{12}$ is now a $6N\times 6N$ matrix where
the decoration now reduces to complete graphs of order $6$ shown in Fig. 2C.

{\bf ii) } The $6N$ $Ext$ entries in $M_a$ are partitioned again into $2$
subsets\ labeling even and odd sites respectively thus exploiting the fact
that $\Sigma $ is a bipartite graph. Schematically $M_a$ can be written as 
\begin{equation}
M_{a=}\left| 
\begin{array}{ll}
M_{EE} & M_{EO} \\ 
M_{OE} & M_{OO}
\end{array}
\right| \; \; .
\end{equation}
Now $M_{EE},M_{OO}$ do not contain $X\,$since an inherited bond always
connects sites of opposite parity while $M_{EO},M_{OE}$ are instead linear
in $X$. We may apply again (\ref{NonFaUnaPiega})\ and obtain a $3N\times 3N$
matrix of the form 
\begin{equation}
M_b(\Phi ,X^2)\;=\;A_b+X^2\;B_b(\Phi )\; \; \;, 
\label{EmmePiega}
\end{equation}
such that $\det(M_a)=\det(M_b)$.

It can be verified that the diagonal matrix $\Omega ={\bf I}_N\otimes
Diag(1,1,1,-1,-1,-1)$ obeys the relations: 
\begin{equation}
\Omega A_c\Omega =A_b^{-1}\;,\;\Omega B_c\Omega =B_b^{-1} \; \; ,
\end{equation}
and that $A_b,B_b$ share the same spectrum with eigenvalues $\pm i,\,\pm
i(2- \sqrt{3})\;,\pm i(2+\sqrt{3})$ independently of $\Phi \;$and $\det
A_b=\det \Omega =1.\,$The essential information is now contained in $B_b\;$
and\ $X$\ since $A_{b\;}$ is a fixed numerical matrix.

By Cayley's theorem the determinant of $M(\Phi ,X)$ or $M_b(\Phi ,X^2)$ is
the square of an even polynomial ${\bf Pf}(\Phi ,X)$ in the activity $X$
called Pfaffian of $M(\Phi ,X)$.

As we shall see, the ambiguity of sign in the extraction of the square root
can be solved by imposing all Pfaffians to be $=1$ in the high\ temperature
limit $X\rightarrow 0$. However the polynomial ${\bf Pf}(\Phi ,X)$ may have
zeros on the real axis in the physical range $X\;=0,1$ and may change sign
when reaching $X=1$. By generalizing the results of ref. \cite{nosotros} we
find that the sign in $X=1$ is given by the function $\sigma ({\bf a)}$
defined below in (\ref{selfinte}), having a precise topological meaning. 

{\bf iii)\ } We have : 
\begin{equation}
\det M_b(\Phi ,X^2) = \det (A_b+X^2\;B_b)=\det ({\bf I} _{3N}+X^2
\;A_b^{-1}B_b)=\det ({\bf I}_{3N}+X^2\;U_b)
\end{equation}
where $U_b =A_b^{-1}B_b$.

Let $\Theta =\Omega A_b\;$so that $\Theta ^2=$ ${\bf I}_{3N}{\bf \;\;}$and 
\begin{equation}
\Theta \,U_b\;\Theta \;=\Theta A_b^{-1}B_b\Theta =\Omega B_b\Omega
A_b=B_b^{-1}A_b=U_b^{-1}  
\label{ThetaU}
\end{equation}

$\Theta \;$is again a block $6\times 6\;$matrix which can be explicitly
diagonalized so that by changing basis both $\Theta \;$and $U\;$can be
written in the following block form where briefly ${\bf I=I}_{12M}={\bf I}
_{3N/2}$ 
\begin{equation}
\Theta =\left| 
\begin{array}{ll}
{\bf I} & 0 \\ 
0 & -{\bf I}
\end{array}
\right| ,U_b=\left| 
\begin{array}{ll}
Q & R \\ 
S & T
\end{array}
\right| \; \; .
\end{equation}
From (\ref{ThetaU}) we see that $Q,R,S,T\;$satisfy the identities 
\begin{equation}
Q^2-RS={\bf I},\;Q=RTR^{-1},Q=S^{-1}TS\;\;.
\end{equation}
We have now

\begin{eqnarray}
&\det& ({\bf I}+X^2\;U_b)= \nonumber \\
&\det &\left| 
\begin{array}{ll}
{\bf I}+X^2\;Q & X^2R \\ 
X^2\;S & {\bf I}+X^2\;T
\end{array}
\right| =\det \left( \left| 
\begin{array}{ll}
{\bf I} & 0 \\ 
0 & S
\end{array}
\right| \left| 
\begin{array}{ll}
{\bf I+}X^2Q & X^2RS \\ 
X^2 & {\bf I+}X^2S^{-1}TS
\end{array}
\right| \left| 
\begin{array}{ll}
{\bf I} & 0 \\ 
0 & S^{-1}
\end{array}
\right| \right) 
\end{eqnarray}
but now $S$ can be included in the changes of basis and we can assume $
S\;=1,\;T\;=Q,R=Q^2-$ ${\bf I}$. Therefore we may write simply  
\begin{eqnarray}
\det ({\bf I}+X^2\;U_b) &=&\det \left| 
\begin{array}{ll}
{\bf I+}X^2Q & X^2(Q^2-1) \\ 
X^2 & {\bf I+}X^2Q
\end{array}
\right| =
 \det \left| 
\begin{array}{ll}
0 & -(X^4+1){\bf I}-2X^2Q \\ 
{\bf I} & \; \; \; \; {\bf I+}X^2Q
\end{array}
\right| \nonumber \\
&=&
\det \left( (X^4+1){\bf I}+2X^2Q\right) =\;X^{24M}\;\det \left(
(X^2+X^{-2}){\bf I}+2Q\right) 
\end{eqnarray}
where we subtracted from the upper block row the lower block row multiplied
by $Q+X^{-2}\;$. In the last form the determinant is evaluated for a matrix $
(X^4+1){\bf I}+2X^2Q$ of rank $12M$, a factor $8$ down from $M(\Phi ,X)$,
with a considerable gain in the computational speed. Clearly ${\bf Pf}(\Phi
,X)X^{-12M}$ is invariant under the map $X\rightarrow X^{-1}$.

\section{Cycles and Co-Cycles over $\Sigma$}

The black elements of $H_1(\Sigma ,{\bf Z_2})$ are generated by two classes
of elementary cycles $E_{i_1}(q),O_{i_1}(q)$, referred to as even or odd
cycles forming the sets $E,O$. Such cycles are homologically equivalent over 
$\Gamma $ to cycles of the form 
\begin{equation}
L_{i_2}+L_{i_3}(D_{i_2}q)+L_{i_2}(D_{i_3}q)+L_{i_3}(q) \; \; ,  \label{EO}
\end{equation}
where $n_{i_1},n_{i_2}$ are even/odd for cycles $\in $ $E$/ $O$

Cycles $\in \{ E,O\}$ are thus generated by the boundaries $\delta F_i(q)$
of the $12M=3N/2$ plaquettes of $\Gamma $ which are not faces of $\Sigma$
(see Fig.1).

However, these cycles are not independent since their number is $
6M>2(1+2M)=2g$. In order to perform a correct counting we consider the
following product of $6$ plaquettes for $n_1+1,n_2,n_3$ even (remind $
q\equiv (n_1,n_2,n_3)$) 
\begin{equation}
E_1(D_1q)+E_1(q)+\delta F_2(q)+\delta F_2(D_2q)+\delta F_3(q)+\delta
F_3(D_3q)\;\;,  \label{cube}
\end{equation}
i.e. the boundaries of the plaquettes of the elementary cube $C(q)$ obtained
by translating of the site $V(q)$ by one unit along the positive direction
of all three axes. 
By expanding the product (\ref{cube}) each bond appears twice,
whence follows the equivalence modulo a boundary of $E_1(D_1q)$ and $E_1(q)$,
which are identified in $E$. Indeed the parity of the last four plaquettes
indicates that they must be faces of $\Sigma$ and are topologically trivial.
Both in $F_3(n_1,n_2,n_3)$,$F_3(n_1,n_2,n_3+1)$ the sum $n_1+n_2$ is odd while
in $F_2(n_1,n_2,n_3),F_2(n_1,n_2+1,n_3)$ the sum $n_1+n_3$ is odd. It
follows that they have boundaries equivalent to the identity of $
H_1(\Sigma ,{\bf Z_2})$ and that 
\begin{eqnarray}
E_i(D_iq) &=&E_i(q)\;,\;\;n_i\;odd  \nonumber \\
O_i(D_iq) &=&O_i(q)\;,\;\;n_i\;even  \; \; .
\label{id1}
\end{eqnarray}

Each cube of $\Lambda $ gives an identity among cycles, but their actual role
depends on the parity of $n_1,n_2,n_3$.

If $n_1,n_2,n_3$ have all the same parity, the boundary of the cube does not
contain faces and leads to the identity 
\begin{equation}
\sum_{i=1}^3\left( E_i(q)+E_i(D_iq)\right) =0\;\;,  \label{id2}
\end{equation}
which connects $6$ cycles of the same parity and must be added to the
previous identities. We shall denote with ${\cal E}$ and ${\cal O}$ the set
of values of $q$ restricted to $n_1,n_2,n_3$ all even and all odd
respectively. If $q\in {\cal E}$, the cubes $C(q)$ and $C(T_iq)$ are
contiguous and contain the common cycle $E_i(D_iq)=E_1(T_iq)$. It follows
that the $M$ identities deriving from the even cubes are not independent
because in their sum each cycle appears twice. The same reasoning applies to
the odd cubes. Thus we may write 
\begin{equation}
\sum_{q\in {\cal E}}E_i(q)=0\;,\;\;\sum_{q\in {\cal O}}O_i(q)=0\;\;.
\label{id3}
\end{equation}

Another class of identities connects even with odd cycles. Consider the
plane of $\Lambda $ with $n_3$ constant. Such a plane is tessellated by
plaquettes which are faces if $n_1+n_2$ is odd and otherwise have boundaries
(on $\Lambda$ but not on $\Sigma$) which are cycles $\in $ $E_3(q)$ or $
O_3(q)$. In virtue of the periodicity of $\Lambda $ the plane has toroidal
topology and the sum of the boundaries of all the plaquettes is 0. We thus
have the identities 
\begin{equation}
e(n_3)=\sum_{n_1,n_2even}E_3(n_1,n_2,n_3)=o(n_3)=
\sum_{n_1,n_2odd}O_3(n_1,n_2,n_3)\;\;.  \label{id4}
\end{equation}

Replacing in the above expressions $n_3$ with $n_3-1$ we obtain the same
identity due to (\ref{id1}).

Such a result can be further extended as follows. If $n_3$ is even, we have $
e(n_3)=e(n_3-1)$. Let's consider the product of the boundaries of the cubes $
C(n_1,n_2,n_3)$ where $n_3$ is a fixed even index and $n_1,n_2$ are
even 
\begin{equation}
\sum_{n_1,n_2 \; {\rm even}, i=1,2,3} \left[
E_i(n_1,n_2,n_3)+E_i(D_i(n_1,n_2,n_3)) \right]\;\;.
\end{equation}
Factors of the type $E_1(n_1,n_2,n_3)$, $E_2(n_1,n_2,n_3)$ always appear
twice in contiguous cubes and the products can be written as 
\begin{equation}
\sum_{n_1,n_2{\rm even}} \left[ E_3(n_1,n_2,n_3)+E_3(n_1,n_2,n_3+1) \right]
=e(n_3)e(n_3+1) \; \;,
\end{equation}
from which we have $e(n_3)=e(n_3+1)$. It follows that $e(n_3)$ does not
depend on $n_3$. The same result holds for all directions and for cycles of
odd arguments. Form (\ref{id4}) we have $e(n_i)=o(n_i)\equiv I_i$, providing
three exceptional identities $I_i$, $i=1..3$ which connect cycles of
different parity.

Finally we have three exceptional cycles $R_i(q)$\ $i=1..3$ given by 
\begin{equation}
R_i(q)=\sum_{n\;=\;0,1,...,N_i-1}L_i(D_i^nq)\;\;.  \label{last}
\end{equation}
We set $R_i=$ $R_i(0)$. Each $R_i(q)$ can be expressed in terms of $R_i$ and
of $E,O$ cycles. Note that $R_i(q)$ does not depend on $n_i$ and hence can
be written in terms of two site labels and a direction, e.g. $
R_1(q)=R_1(n_2,n_3)$. Geometrically $R_1(q)$ is a straight circuit which
winds around $\Lambda $ by exploiting the periodicity, its length is $
N_1=2M_1$ and is even (see Fig. 1). There is no ambiguity in extending
formula (\ref{circuit}) to black $R_i(q)$. The existence of $R_i(q)$ follows
from the periodicity conditions imposed on $\Lambda $ and also exists in the 
$2D$ Ising model where they produce the analog of $R_i(q)$ for $i=1,2$. There
is however no analog in $2D$ of the $E,O$ cycles.

We are now in position to check that the overall number of independent
cycles is in fact $2g$. To each one of the $M$ even/odd cubes corresponds
one identity but only $M-1$ of them are independent. The $3M$ even/odd
cycles we started with reduce to $3M-(M-1)=2M+1=g$ independent ones, giving
exactly an overall number of $2g$. The three cycles that are eliminated by
the identities $I_i$ are replaced by the exceptional cycles $R_i$ so that
the overall counting remains unaltered.
(In the thermodynamic limit $N \to \infty$, we expect a negligible 
contribution from the surface terms $R_i$).

We can restate the counting problem by noticing that the total number of
defined cycles amounts to $3 M+3 M+3=6 M+3=3 g$. However the total number of
identities is given by $2 (M-1)+3=g$ and hence the number of independent
cycles is once more $4 M+2=2 g$ as expected.

The generalized dimer method amounts to writing the partition function as a
sum of $2^{2g}$ $G$-invariant Pfaffians associated to all possible different
orientations $\Phi$.

We call {\bf fundamental} orientation $\Phi _F$ that for which $\Phi ({\bf a}
)=1$ for all ${\bf a\;\in }H_1(\Sigma ,{\bf Z_2})$. This can be obtained by
setting 
\begin{equation}
\phi _i(n_1,n_2,n_3)=(-1)^{n_i} \; \;.  \label{Fund}
\end{equation}

While the fundamental orientation makes no distinction among plaquettes, the 
{\bf antifundamental} $\Phi _A$ requires $\Phi _i(q)=-1$ for all $E,O$
cycles but $\Phi (R_i)=1$. There is no simple recipe for $\Phi _A$ analogous to
(\ref {Fund}) since the way it appears depends on the spanning tree ${\cal T}$.

\section{Topological Intersections and Sign Functionals}

\subsection{Intersection of cycles}

The bilinear symmetric functional $I[{\bf a},{\bf b}]$ over the cycles ${\bf 
a},{\bf b}$ is given by ${\rm Mod}(p,2)$ where $p$ is the number of
intersections of ${\bf a},{\bf b}$ over $\Sigma$. In our case a detailed
analysis shows that cycles $\in E,O$ of same parity or same normal do not
intersect. The general formula can be deduced from 
\begin{equation}
I[E_1(m_1,m_2,m_3),O_2(n_1,n_2,n_3)]=\delta _{n_1,m_1-1}\delta
_{m_2,n_2-1}\delta _{n_3,m_3+1}\;\;,  \label{start}
\end{equation}
by imposing invariance under cyclic permutation of the axes and under the $
S,T_i$ symmetry operations. Two adjacent cycles do not intersect.

The cycles $R_i$ do not obey this rule and the intersections with the exceptional cycles are given by
\begin{eqnarray}
I[R_k &,&E_i(q)]=0\;\;{\rm if}\;i\neq k  \nonumber \\
I[R_k &,&O_i(q)]=0\;\;{\rm if}\;i\neq k  \nonumber \\
I[R_1 &,&E_i(n_1,n_2,n_3)]=\delta _{i,1}(\delta _{n_2,0}\delta
_{n_3,0}+\delta _{n_2,N_2-1}\delta _{n_3,N_3-1})  \nonumber \\
I[R_1 &,&E_i(n_1,n_2,n_3)]=\delta _{i,2}(\delta _{n_1,0}\delta
_{n_3,0}+\delta _{n_1,N_1-1}\delta _{n_3,N_3-1})  \nonumber \\
I[R_1 &,&E_i(n_1,n_2,n_3)]=\delta _{i,3}(\delta _{n_1,0}\delta
_{n_2,0}+\delta _{n_1,N_1-1}\delta _{n_2,N_2-1})\;\;\;,  \label{special}
\end{eqnarray}
together with the expressions obtained by interchanging $E$ with $O$.

The definition of $I[{\bf a},{\bf b}]$ can be extended by linearity\ mod $2$
to arbitrary pairs of cycles ${\bf a},{\bf b}$.

\subsection{Topological excitation and signature}
The intersection formalism leads naturally to the notion of elementary
topological excitations $\{\tau _i(q)\}$, a set of minimal operators which
change locally a proper orientation into a new one, still proper but
inequivalent. $\tau$ acts on the inherited bonds only and we can define $
\tau $ directly over $\Sigma$. Any two inequivalent proper orientations are
connected by a sequence of $\tau$ operations and hence $\{\tau _i(q)\}$
generate all orientations of $\Sigma $. In particular we may reach any
orientation by repeatedly applying $\tau _i(q)$ to $\Phi _0$. In the
following we use the equivalent notations $\Phi _i(n_1,n_2,n_3)\equiv \Phi
_i(q)\equiv \Phi ({\bf a})$ where ${\bf a}=E_i(q)$ or $O_i(q)$. The
elementary topological excitations are in one to one correspondence with the 
$E,O$ cycles and we may adopt the same set of indices and distinguish
between even or odd excitations, $\tau _i(q)\to \tau _i^E(q),\tau _i^O(q)$.

The action of $\tau _1^E(n_1,n_2,n_3)$ $=\tau _1^E(n_1-1,n_2,n_3)$ is given
by 
\begin{equation}
\phi _1(n_1- 1,n_2+\epsilon,n_3+\xi) \to -\phi _1(n_1-1,n_2+\epsilon,n_3+\xi)
\; \; \; \epsilon,\xi=0,1 \; \;.
\label{bomb}
\end{equation}
The definition extends in an obvious way to the other directions and, in
virtue of the invariance under the $S,T_i$ operations, (\ref{bomb}) hold also
for odd operators. Clearly, the values of $\phi$ on bonds not appearing in ( 
\ref{bomb}) are unaffected by $\tau _1^E(q)$.

Given an elementary cycle, e.g. $E_1(q)$, there exist four other
intersecting cycles, e.g. 
\begin{eqnarray}
&&O_2(D_1^{-1}D_2D_3^{-1}q)\;,\;\;O_2(D_1^{-1}D_2D_3q)\;,  \nonumber \\
&&O_3(D_1^{-1}D_2^{-1}D_3q)\;,\;\;O_3(D_1^{-1}D_2D_3q)\;\;.  \label{intcycle}
\end{eqnarray}

Consider now the orientation $\Phi _2(n_1-1,n_2+1,n_3-1)$ corresponding to $
O_2(n_1-1,n_2+1,n_3-1)=O_2(D_1^{-1}D_2D_3^{-1}q)$,

\begin{eqnarray}
\Phi_2(n_1-1,n_2+1,n_3-1)& =& \phi_1(n_1-1,n_2+1,n_3-1) 
\phi_3(n_1,n_2+1,n_3-1) \times  \nonumber \\
& & \phi_{-1}(n_1,n_2+1,n_3) \phi_{-3}(n_1-1,n_2+1,n_3) \; \;.
\label{example}
\end{eqnarray}

The last factor changes sign under the action of $\tau _1^E(q)$ and hence
the orientation of $\Phi _2(n_1-1,n_2+1,n_3-1)$ also changes and the same
result holds for the other cycles of the example (\ref{intcycle}).

The action of $\tau _i^{E,O}(q)$ changes $\Phi $ into an inequivalent
orientation which differs from $\Phi $ only in the orientations of the local
cycles intersecting $E_i(q),O_i(q)$. The orientation of all faces of $\Sigma$
remain unchanged under $\tau _i^{E,O}(q)$ so that the Kasteleyn's conditions
are always fulfilled. Each cycle can be given an active role if identified
with $\tau _i^{E,O}(q)$ or a passive one if considered as a cycle changing
parity under the action of $\tau _i^{E,O}(q)$. The functional ${\cal I}_{
{\bf a}}\in H^1(\Sigma ,{\bf Z_2}):\;H_1(\Sigma , {\bf Z_2})\rightarrow {\bf 
Z_2}$ is defined by 
\begin{equation}
{\cal I}_{{\bf a}}({\bf b})\;=\;(-1)^{I[{\bf a},{\bf b}]}\;,\;{\bf a}, {\bf 
b }\in H_1(\Sigma ,{\bf Z_2}).  \label{Ifunctional}
\end{equation}

\subsection{Axial gauge}
We fix the gauge for the orientations of bond by selecting a subset of $N-1$
bonds forming a spanning tree ${\cal T}$ of $\Lambda $, that contains all
the sites of $\Lambda$. A convenient gauge fixing is the Axial Gauge (as
displayed in Fig. 4) 
\begin{eqnarray}
&\phi& _3(0,0,n_3) =1 \; \; \; \; n_3=0,1,...,N_3-2  \nonumber \\
&\phi& _2(0,n_2,n_3)
=(-1)^{n_3}, \; \;\; \;  n_2=0,1,...,N_2-2, \; \; \; \;
n_3=0,1,...,N_3-1  \nonumber \\
&\phi& _1(n_1,n_2,n_3)
=(-1)^{n_1}, \; \; \; \; n_1=0,1,...,N_1-2, \; \; \; \; n_2=0,1,...,N_2-1,
\; \; \; \; n_3=0,1,...,N_3 \;.  
\label{symmgauge}
\end{eqnarray}

The axial gauge leaves undetermined the orientations of $3N-(N-1)=2N+1=16M+1$
bonds not belonging to ${\cal T}$ . However, the $12M-1$ Kasteleyn's
conditions reduce the number of such independent orientations to $
16M+1-12M+1=4M+2=2g$, as expected.

From the recursive relation equivalent to (\ref{twocircuits}) 
\begin{equation}
\Phi ({\bf a}+{\bf b})=\Phi ({\bf a})\Phi ({\bf b})(-1)^{I[{\bf a},{\bf b}
]}\;\;,  \label{recursion}
\end{equation}
we deduce then the value of $\Phi$ on all the elements of 
$H_1(\Sigma ,{\bf Z_2})$ starting from $\Phi _i(q)$. 
From (\ref{twocircuits}) we see that $\Phi _i(q)$ is
invariant under ${\cal G}$ and the same is therefore true for all $\Phi ( 
{\bf a})$. The set of all the $\Phi _i(q)$ determines therefore the global
orientation $\Phi$ (mod ${\cal G}$) of $\Sigma $.

\section{General procedure and Pfaffian expansion over $\Sigma$}

The geometrical structure on which we define Pfaffians admits an alternative
equivalent definition. We consider $2$ cubic lattices $\Lambda_E,\Lambda_O$
each having $M$ vertices in $1-1$ correspondence with even/odd cubes $
C(n_1,n_2,n_3)$ where $n_1+n_2+n_3=$ even/odd, respectively. The bonds of
the lattices correspond to the even/odd $E,O$ cycles.

In place of spins we have orientation parity of the $E,O$ cycles, satisfying
the identities (\ref{id1}--\ref{id4}). Topology plays a marginal role in the
classical $2D$ Ising lattice where the genus $g=1$ leads to a sum over 4
pfaffians only.

Here the reduction of the original sum on $N=4(g-1)$ spins to one on $2g$
functionals does not solve the problem but reduces the complexity of the
task. The orientations on the sublattices $\Lambda _E,\Lambda _O$ are not
independent.

We sketch now the general algorithm which fixes all gauges and gives all
bond orientations $\phi _i(q)$ in terms of a subset of $2g$ independent $
\phi ^{\prime }s$ by means of the Kasteleyn's conditions. We consider anly the
homogeneous case where the activities of bonds are the same in all
direction, however the results can be straightforwardly generalized to any
non homogeneous distribution of bond interaction energies.

The actual steps are quite complex and can be summarized as follows:

\noindent
{\bf (1.)} Compile a list of all $3N$ $\phi_i(q)$ considered as independent
binary variables.

\noindent
{\bf (2.)} Impose the axial gauge (\ref{symmgauge}) and obtain a sublist of $
2N+1$ terms $\phi _i(q)$ only.

\noindent
{\bf (3.)} Impose the Kasteleyn conditions over faces. This can be achieved by
solving iteratively (\ref{Kast}) thus reducing the number of independent $
\phi _i(q)$ to $2g =2+N/2$ forming a basis ${\cal B}$ for the $2^{2g}$
functionals in $H^1(\Sigma ,{\bf Z_2})$. From ${\cal B}$ we define by group
multiplication the generic functional 
$\Phi ({\bf a)} \in H^1(\Sigma ,{\bf Z_2})$, $ \forall {\bf a}$ 
${\bf \in }H_1(\Sigma ,{\bf Z_2})$.

\noindent
{\bf (4.)} Select a complete basis $\Omega =\{{\bf \omega }_k\;,k=1..2g\}$ of
independent cycles out of $E,O,R_1,R_2,R_3$. A generic cycle can then be
written as 
\begin{equation}
{\bf a}=\sum_{k=1}^{2g}\;e_k\;{\bf \omega }_k\;\;,e_k=0,1  \label{baseomega}
\end{equation}

\noindent
{\bf (5.)} Define the self intersection function 
$H_1(\Sigma ,{\bf Z_2})\rightarrow {\bf Z}_2$,
see (\ref{start}), (\ref{special}), (\ref{recursion}) 
\begin{equation}
\sigma ({\bf a)}=\;(-1)^{\sum_{k=2}^{2g}\sum_{k^{\prime }=1}^{k-1}I[{\bf 
\omega }_k, {\bf \omega }_{k^{\prime }}]\;e_ke_{k^{\prime }}}
\label{selfinte}
\end{equation}

\noindent
{\bf (6.)} The function ${\cal I}_{{\bf a}}$ in (\ref{Ifunctional}) defines an
invertible duality map $H_1(\Sigma ,{\bf Z_2})\rightarrow H^1(\Sigma ,{\bf
Z_2})$. The map $\sigma \circ $ ${\cal I }_{{\bf a}}^{-1}$ lifts the self
intersection to $H^1(\Sigma ,{\bf Z_2})$. In any case we do not need to
compute explicitly the inverse of ${\cal I}_{{\bf a}}$.

\noindent
{\bf (7.)} Given ${\bf a}$ as in (\ref{baseomega}), we compute ${\cal I}_{{\bf 
a }}$ and expand it in terms of the restricted basis ${\cal B}$ defined in
step {\bf 3} , thus determining the orientations $\Phi ({\bf a),}$ $
\phi_i(q) $ and all matrix elements of $M(\Phi ({\bf a))}$ explicitly as
functions of $e_1,...,e_{2g}$.

\noindent
{\bf (8.)} The dimer generating function ${\bf Z}_0(X)$ is then given by the
sum over Pfaffians 
\begin{eqnarray}
{\bf Z}_0(X)&=&\frac 1{2^g}\sum_{{\bf a\in }H_1}\sigma ({\bf a}){\bf Pf}
(\Phi ({\bf a}),X)  \nonumber \\
&=& \frac 1{2^g}\sum_{\{{\bf e}_k=\,\,0,1\}} (-1)^{\sum_{k=1}^{2g}
\sum_{k^{\prime }=1}^{k-1}I[{\bf \omega }_k, {\bf \omega }_{k^{\prime }}]\;
e_ke_{k^{\prime }}}\,\,\, {\bf Pf}(\Phi(\sum_{k=1}^{2g}\;e_k\;{\bf \omega }
_k),X) \; .  \label{SommaZeta}
\end{eqnarray}

As in eq.(\ref{generating}), the Ising partition function is simply given by 
\begin{equation}
Z=(2\cosh (\beta J))^{3 N } {\bf Z}_0(X) \; \;,  
\label{Z3D}
\end{equation}
where $J$ is the spin-spin interaction energy and $X=\tanh(\beta J) $ is the
activity of a bond at inverse temperature $\beta$.

\noindent
{\bf (9.)} For non-bipartite lattices like the decorated spin lattices there
is no direct and fast numerical method to compute Pfaffians with their
proper sign. This is possible in other cases, for example in dimer coverings
of bipartite lattices where the matrix $M(\Phi ({\bf a))}$ is block
diagonal. In spin lattices we first compute the determinant of $M$, extract
the positive root and set ${\bf Pf}(\Phi ,0)$ positive. The sign at $X=1$ is
then directly given by $\sigma ({\bf a)}$ or obtained by analytic
continuation of the Pfaffian which is a polynomial of degree $3N$ in $X$.
Thus $\sigma ({\bf a)}$ could be used to predict the parity of the number of
real zeros of the Pfaffian in the interval $0<X<1$.

\section{Preliminary analysis of Pfaffians}

This section is very preliminary and deals with a number of properties and
conjectures which are essential in analyzing the behavior of the Pfaffian
expansion. Some of these properties have been verified on finite Ising
lattices and are in agreement with extensive numerical sampling and
numerical findings \cite{zeros}. We should mention that the procedure has
not yet been optimized for a numerical approach. In particular the
classification of Pfaffian symmetries has still to be implemented.

The self-intersection function $\sigma ({\bf a)}$ can be defined on any
triangulated surface of genus $g$ and particular examples of genus $g=0$ to 
$3$ have been worked out in detail in the literature\cite{McCoy,nosotros}. 
The argument ${\bf a}$ takes $2^{2g}$ values parametrized by $2g$ binary
variables $e_1,...,e_{2g}$ . It is always possible to redefine the basis in
$H_1(\Sigma ,{\bf Z_2})$ in such a way as to have $\sigma ({\bf
a)}=\;(-1)^{\sum_{i=1}^g\,e_i\,e_{i+g}}= \prod_{i=1}^g\;(-1)^{e_i\;e_{i+g}}$.
The factor $(-1)^{e_i\;e_{i+g}}$ takes $3$ times the value $1$ and once the
value $-1$ as $e_i,e_{i+g}$ run on $ 0,1$ and all the factors appearing in
$\sigma ({\bf a)}$ are independent. Suppose now that $N_{+}(g)$, $N_{-}(g)$
are the number of times $\sigma ( {\bf a)}$ takes the value $1,-1$
respectively so that $N_{+}(1)=3$ and $ N_{-}(1)=1$. We have the recursion
relation
\begin{equation}
N_{+}(g+1) = 3\,N_{+}(g)+ N_{-}(g) \;, \; \; \; 
N_{-}(g+1) = N_{+}(g)+3 N_{-}(g) \; \;,
\end{equation}
which has the solution 
\begin{equation}
N_{+}(g) = 2^{g-1}(2^g+1) \;, \; \; \;  
N_{-}(g) = 2^{g-1}(2^g-1) \; \; .  
\label{PiuMeno}
\end{equation}
$N_{+}(g)$ and $N_{-}(g)$ give the total number of positive and negative $
\sigma ({\bf a)}$ in the expansion (\ref{SommaZeta}). This statistics is
important in evaluating the convergence properties of the expansion. For
high $g$ this means that if we pair off positive and negative $\sigma ({\bf 
a)}$ we are left with a small excess of $2^g$ positive values, i.e. one part
in $2^g$. Since ${\bf Pf}(\Phi ,0)=1$ we have ${\bf Z}_0(0)=1$ as expected
at $T=\infty$. As $X$ increases all terms in (\ref{SommaZeta}) become
eventually all positive and equal to ${\bf Pf}(\Phi_F,1)$. Comparing the
total sum $2^{2g}{\bf Pf}(\Phi _F,1)$ with the known $T\rightarrow 0$ limit
for ${\bf Z}_0$ we obtain ${\bf Pf}(\Phi _F,1)=2^{14M}$. The equality of the
absolute values of ${\bf Pf}(\Phi ,1)$ hold only on decorated spin lattices
whereas generic dimer lattices have a spectrum of values. The factor $\sigma
({\bf a})$ induces a cancellation with a cutoff factor $2^{-g}$ in (\ref
{SommaZeta}) in the $T\rightarrow \infty $ or $X\rightarrow 0$ limit which
does not occur at $T\rightarrow 0$ and leads to a steeper log derivative for 
${\bf Z}_0(X)$ as compared to that of the single Pfaffians.

As first noted by Wannier \& Kramer \cite{WK} planar spin lattices can be
characterized by duality relations. However for $\ g>0$ duality does not
relate directly the partition function of a lattice to that of the dual but
rather acts very simply on the single terms in the expansion (\ref{SommaZeta}
) by changing their signs only so that they are still positive in the $T$ $
\rightarrow \infty$ limit. Since duality swaps $T\rightarrow \infty $ and $
T\rightarrow 0$ limits we see that in dual lattices positivity is required
at opposite ends of the interval $0<X<1$. In the $2D$ Ising lattice the sign
reversal does not alter the partition function in the thermodynamical limit
but the same need not to be true in $3$ dimensions. In any case the $3D$
lattice is not self dual.

A lower bound\ $X_0$ for the zeros of ${\bf Pf}(\Phi ,X)$ can be derived
from (\ref{EmmePiega}). From the spectrum of $A_f,B_f$ we get $\left\|
A_f\right\| \;<2+\sqrt{3},\left\| B_f\right\| \;<2+\sqrt{3}$ whence\ $
X_0\;>\;\frac 1{2+\sqrt{3}}=2-\sqrt{3}\;\simeq \;0.267...$. Numerical
analysis indicates that in fact Pfaffians vanish in the range $
X_{m}=0.3178...$\ which is actually reached by \ ${\bf Pf}(\Phi _F,X)$ to 
$X_{M}=\;0.3506...$ reached by ${\bf Pf}(\Phi _A,X)$. These values can be
computed exactly as roots of an algebraic equation because the translational
symmetry of these Pfaffians leads to explicit formulas . Random sampling of
orientations $\Phi$ up to $N_1=N_2=N_3=8$ indicate that zeros tend to
accumulate around $0.34$, see Fig. 5.

Therefore, we divide now the interval $0<X<1$ into $4$ regions :

\begin{itemize}
\item[(I)]  The very low temperature region $X>X_{M}$ where all terms in
the expansion (\ref{SommaZeta})\ of ${\bf Z}_0$ are positive and the series
converges rapidly and agrees with low $T$ expansions.

\item[(II)]  The crossover region centered in $X=0.34$ where $
N_{-}(g)=2^{g-1}(2^g-1)$ Pfaffians change sign.

\item[(III)]  The region between the estimated critical temperature $
X=X_0\simeq 0.2108$ \cite{zeros} and $X_{m}$. In this region the negative
terms have absolute values smaller that the positive terms and we expect
that convergence degrades rapidly as we move toward $X_0.$

\item[(IV)]  High temperature region $X<X_0$ where all absolute values of
the terms in (\ref{SommaZeta}) becomes comparable and the value of ${\bf Z}
_0 $ is determined by $2^g$ unpaired Pfaffians. For small $X$ all absolute
values of the Pfaffians are close to $1$ and the sum (\ref{SommaZeta}) over a
sample of $L$ random terms has a noise of the order of $2^{-g}\sqrt{L}$
while we expect a signal of the order of $2^{-2g}L$, normalized to $1$
at $X=0$ for the complete sum. In order to get a signal we need a ratio
signal/ noise $\frac{2^{-2g}L}{2^{-g}\sqrt{L}}\simeq 1$ i.e. $L\simeq 2^{2g}$.
This means that unless one sums over all terms we get only noise and that
numerical computation is ruled out unless one obtains an explicitly summable
formula for the Pfaffians as it happens in the limit $T\rightarrow \infty $.
Therefore exact matching with known $T\rightarrow \infty $\ results is still
possible and useful. As $X$ increases the absolute values of positive terms
grows on the average more rapidly than that of the negative ones thus
improving the signal/noise ratio. We conjecture that the critical $X=$\ $X_0$
is effectively a threshold beyond which the signal becomes effective. In
general for a fixed $X$ and all $\Phi $ we have ${\bf Pf}(\Phi _A,X)\leq 
{\bf Pf}(\Phi ,X)\leq {\bf Pf}(\Phi _F,X)$.
\end{itemize}

Details and high/low temperature expansion will be discussed in a forthcoming
paper\cite{preparation}.

\section{Dimer Statistics}

A simple application of the above formalism is the evaluations of the number
of Perfect Matching, i.e. dimer coverings, over the 3D cubic lattice $
\Lambda $.

As for the Ising model, it can be solved exactly or can be
treated easily in the case of planar lattices whereas it still 
represents an open problem in the case of non-planar graphs \cite{dimers}.

The dimer covering generating function is given by a Pfaffian expansion
similar to (\ref{SommaZeta}) where now $M(\Phi )$ is a $N\times N$ matrix of
elements $M_{q,q^{\prime }}(\Phi )=\phi _i(q)=\pm 1$ depending on the
orientation of the bond. The decorating bonds are absent whereas the
orientations $\Phi $ of $\Sigma $ play exactly the same role as in the Ising
case. Following the same steps discussed for the Ising case and separating
odd and even sites, we arrive to a block diagonal form of $M(\Phi )$

\begin{equation}
M(\Phi )=\left( 
\begin{array}{ll}
{\bf 0} & C(\Phi ) \\ 
-C(\Phi )^T & {\bf 0}
\end{array}
\right)
\end{equation}

Now we have directly ${\bf Pf}(M(\Phi ))=Det(C(\Phi ))$, and the expansion
reads:

\begin{eqnarray}
{\bf Z}_{Dimers} &=&\frac 1{2^g}\sum_{{\bf a\in }H_1}\sigma ({\bf a}) {\bf 
Pf }(\Phi ({\bf a}))  \nonumber \\
&=& \frac 1{2^g}\sum_{\{{\bf e}_k=\,\,0,1\}}
(-1)^{\sum_{k=2}^{2g}\sum_{k^{\prime }=1}^{k-1}I[{\bf \omega }_k, {\bf 
\omega }_{k^{\prime }}]\;e_ke_{k^{\prime }}}\,\,\, {\bf Det}(C\left( \Phi
\right)),\,\,\,\,\; \Phi \;\equiv \Phi (\sum_{k=1}^{2g}\;e_k\;{\bf \omega }
_k)  \label{dimersum}
\end{eqnarray}

Such a formula can be used both for exhaustive enumerations of coverings $
h(N_1,N_2,N_3)$ of finite lattices of linear size $N_1\times N_2\times N_3$
as well as in a probabilistic framework \cite{preparation}.

We have applied (\ref{dimersum}) to the case of finite cubic lattices with
open boundaries in order to recover and improve the known results. The
limitations in the size arise from the number of terms appearing in the
expansion which increase exponentially with the genus of the surface, which
for open boundaries grows as $
g=M_1M_2(M_3-1)+M_2M_3(M_1-1)+M_3M_1(M_2-1)-M_1M_2M_3+1$ ($g=2 L^3-3 L^2+1$
in the isotropic case of linear size $L$). We have found $
h(4,4,4)=5051532105 $ (in agreement with \cite{mathconst}), $
h(6,4,4)=932814464901633$ and $h(6,6,4)= 123115692449982216049513$.

Note that the rigorous lower bound \cite{Hammersley} to the number of dimer
coverings in 3D in the $M\to \infty$ limit can be easily recovered in our
approach by computing, via Fourier Transform, the periodic Pfaffian
corresponding to $\Phi_F$ \cite{preparation}.

Eq. (\ref{dimersum}) can be also thought of as the expansion in terms of
determinants of the permanent of $0-1$ matrices, a $\#P-complete$ problem
which can be easily mapped onto the evaluation of dimer coverings over an
associated bipartite lattice \cite{Minc}.

\section{Conclusion}

In this paper we propose a combinatorial/topological formalism for the study
of the Ising problem over lattices of arbitrarily high topological genus
which generalizes the well-known approach of Kasteleyn.
The partition function is written as a sum over Pfaffians with a 
topological signature. 

We apply the method to the $3D$ cubic Ising problem where we have reached a
very preliminary assessment on the expansion in the high and low temperature
ranges. 
The same formalism applies to the perfect matching problem and provides a 
determinant expansion for the permanent of 0-1 matrices.

Work is in progress on the physical and algorithmic relevance of the
method.

\noindent
{\bf Acknowledgments}

We thank A. Ceresole, M. Rasetti, YU Lu, F. Ricci-Tersenghi, S. Moroni 
for discussions.

\begin{figure}
\centerline{\psfig{file=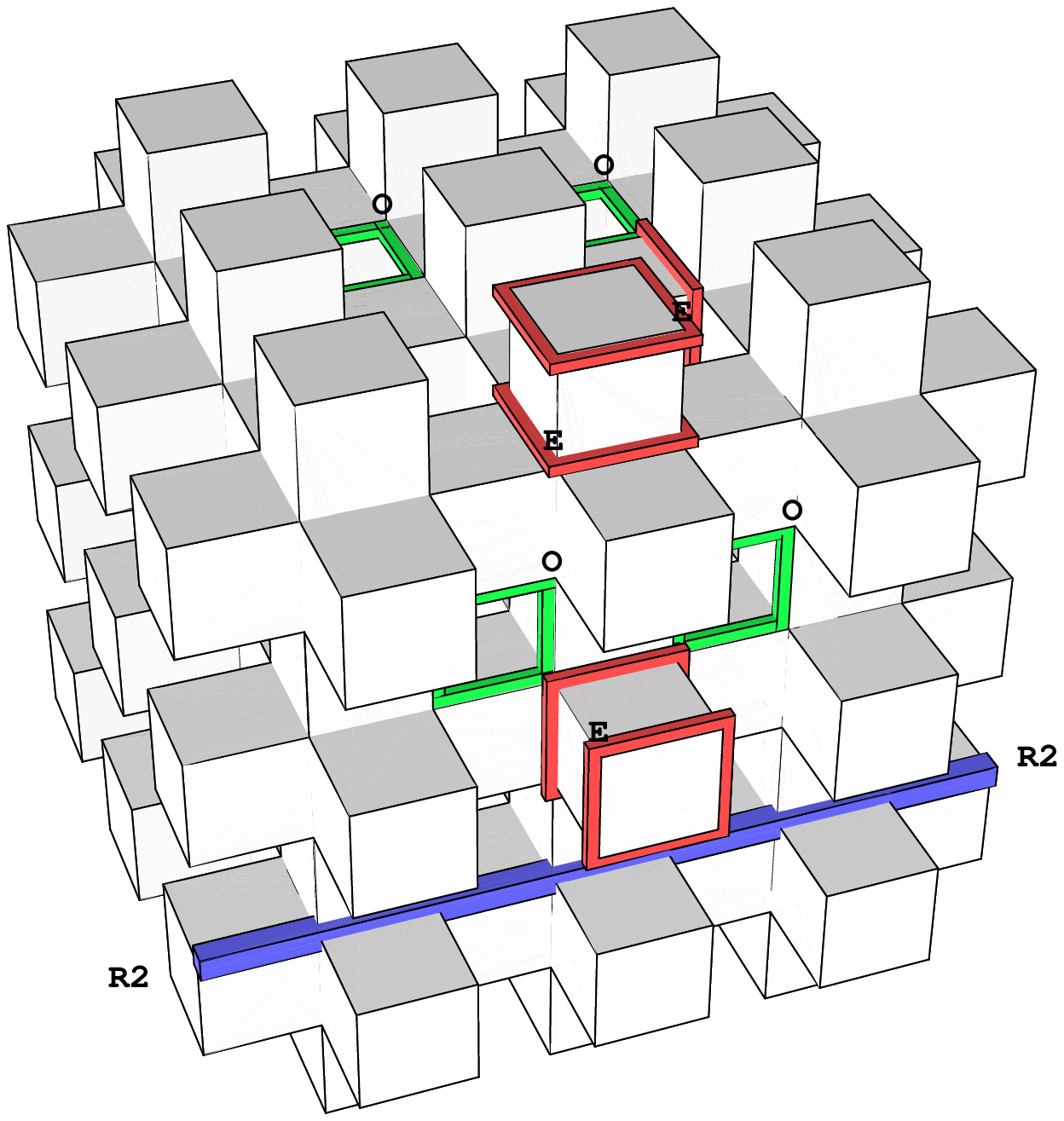,width=12cm}}
\caption{Orientable surface $\Sigma$ of genus $g=N/4+1$ containing all sites
and bonds of the 3D cubic lattice $\Lambda$. Some examples of the
even (E), odd (O) and exceptional (R) cycles are also shown.}
\label{fig1}
\end{figure}

\vfill
\eject

\vspace{3cm}

\begin{figure}
\centerline{\psfig{file=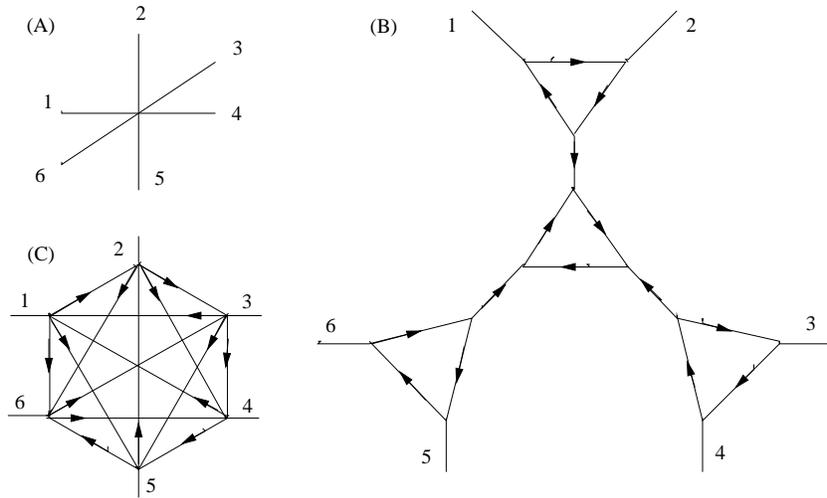,width=11cm}}
\caption{Oriented decorating Clusters. Each site (A) is replaced by
either Fisher's planar decoration (B) or the equivalent complete graph (C).}
\label{fig2}
\end{figure}

\vspace{3cm}

\begin{figure}
\centerline{\psfig{file=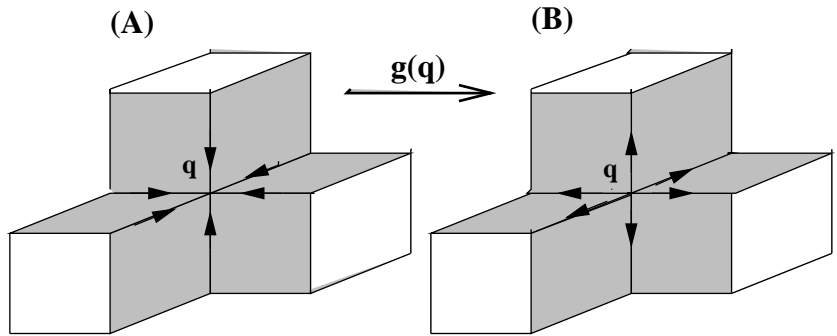,width=11cm}}
\caption{Elementary {\it gauge} operation. The orientation parity of faces
remains unaltered.}
\label{fig3}
\end{figure}

\begin{figure}
\centerline{\psfig{file=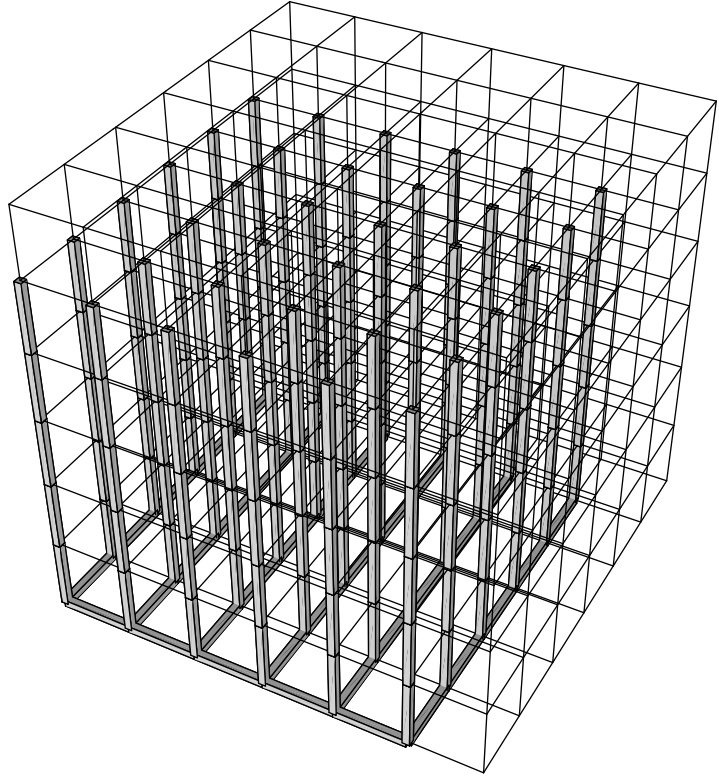,width=10cm}}
\caption{ 3D simple cubic lattice and the Axial Gauge Tree ${\cal T}$}
\label{fig4}
\end{figure}

\begin{figure}
\centerline{\psfig{file=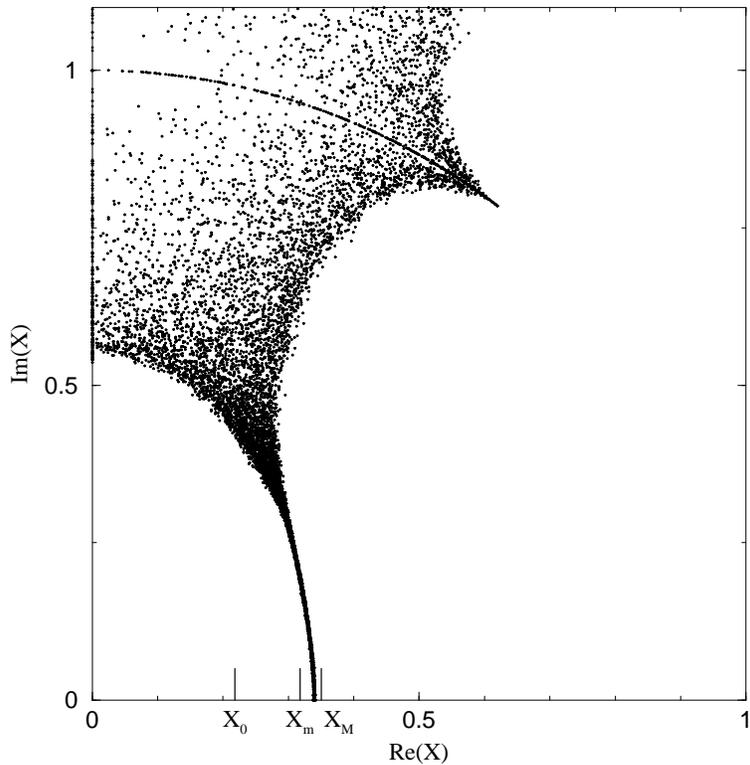,width=10cm}}
\caption{Overlap of the complex zeros of 50 Pfaffians corresponding to
different randomly chosen orientations ($N_1=N_2=N_3=8$). 
$X_M$ and $X_m$ (computed in the $N\to \infty$ limit)
are the upper ond lower bounds for the zeros
given by the singularities of the fundamental and antifundamental Pfaffians
respectively. $X_0$ corresponds to the value of the critical temperature 
estimated by different analytical and numerical methods.}
\label{fig5}
\end{figure}

\end{document}